\documentclass[preprint2]{aastex}

\shorttitle{Variations of subhalo abundance}
\shortauthors{Ishiyama et al.}

\begin{document}


\title{Variation of the Subhalo Abundance in Dark Matter Halos}


\author{TOMOAKI \textsc{Ishiyama}\altaffilmark{1,2}, 
TOSHIYUKI \textsc{Fukushige}\altaffilmark{3}, and
JUNICHIRO \textsc{Makino}\altaffilmark{1}}
\altaffiltext{1}{National Astronomical Observatory, Mitaka, Tokyo 181-8588, Japan ;
ishiyama@cfca.jp, makino@cfca.jp}
\altaffiltext{2}{Department of General System Studies, College of Arts and Sciences,
University of Tokyo, Tokyo 153-8902, Japan}
\altaffiltext{3}{K\&F Computing Research Co., Chofu, Tokyo 182-0026, Japan, fukushig@kfcr.jp}




\begin{abstract}
We analyzed the statistics of subhalo abundance of galaxy-sized and
giant-galaxy-sized halos formed in a high-resolution cosmological
simulation of a 46.5Mpc cube with the uniform mass resolution of
$10^6 M_{\odot}$. We analyzed all halos with mass more than $1.5 \times
10^{12}M_{\odot}$ formed in this simulation box. The total number of
halos was 125. We found that the subhalo abundance, measured by the
number of subhalos with maximum rotation velocity larger than 10\% of
that of the parent halo, show large halo-to-halo variations. The
results of recent ultra-high-resolution runs fall within the variation
of our samples. We found that the concentration parameter and the
radius at the moment of the maximum expansion shows fairly tight
correlation with the subhalo abundance. This correlation suggests that
the variation of the subhalo abundance is at least partly due to the
difference in the formation history. Halos formed earlier have smaller
number of subhalos at present.

\end{abstract}





\section{Introduction}

The cold dark matter (CDM) model is widely accepted as the standard
theory of the formation and evolution of the universe.  According to
this CDM model, the structure formation in the universe proceeds
hierarchically,  from small-scale structures to larger-scale ones.

The CDM model successfully reproduced large-scale structures 
\citep[e.g.,][]{Davis1985, Springel2005}.  However, it has been claimed that
there are serious discrepancies between the structure predicted by
the CDM model and that observed, in the scale of galaxy-sized objects
and below.

\citet{Moore1999a} and \citet{Klypin1999} calculated the evolution
of galaxy-scale dark mater halos using high-resolution cosmological
$N$-body simulations.  They found that dark matter halos of the mass
comparable to the Local Group contained far too many subhalos compared
to the number of known dwarf galaxies in the Local Group. In the case
of dark matter halos of the size of typical clusters of galaxies, the
theoretical prediction and observation agreed pretty well. The
theoretical prediction for galaxy-sized and cluster-sized halos are
similar, and observations of clusters of galaxies and that of Local
Group are very different.

This discrepancy, now called ``missing dwarf problem'', was studied by
many researchers. Follow-up simulation studies 
\citep[e.g.,][]{Diemand2004, Reed2005, Kase2007} gave essentially the same results
as those of \citet{Moore1999a} and \citet{Klypin1999}. 
Accordingly, it is now regarded as one of the most
serious problems of the CDM model.

A number of solutions for this problem have been proposed.  For
example, we can reduce the number of subhalos in small scales by
changing the nature of dark matter in small scales. Models in this
direction are warm dark matter \citep{Kamionkowski2000} and
self-interacting dark matter \citep{Spergel2000}.  Another
direction is to explain the discrepancy as the difference between the
number of nonobserved dark matter subhalos and that of observed
satellite galaxies. Maybe not all subhalos host dwarf galaxies, or the
measured velocity dispersion of dwarf galaxies might be much smaller
than the rotation velocity of the parent subhalos. Models in this
direction includes the suppression of star formation by early
reionization \citep{Susa2004} and self-regulation of star
formation in small halos \citep{Stoehr2002, Kravtsov2004}.
However, none of these explanations are widely accepted as the
clear-cut solution for the missing dwarf problem.

In almost all previous studies of substructures in dark-matter
halos, the so-called re-simulation method has been used.  In this
method, one simulation is done in the following two steps. In the
first step, a relatively large volume (for galaxy-sized halos
typically a 50-100Mpc cube) is simulated with low-mass resolution, and
the candidate regions for high-resolution simulations are identified.
In the second step, the selected regions are simulated with higher
mass resolution. In practically all recent high-resolution
simulations, this re-simulation method is used
\citep{Diemand2007, Diemand2008, Stadel2008, Springel2008}.

This re-simulation method allows us to resolve the structure of a
selected halo with a very high resolution.  On the other hand, it is
not clear what kind of selection bias is introduced by the use of this
method. Typically, the authors claimed that the halo
which looks similar to the halo of the Local Group was selected to
compare with the Local Group. However, in most cases the selection
criteria were not described in detail, and thus it is difficult to see
if any selection bias affected the results or not.

An obvious way to avoid the selection bias, which might be introduced
by the use of the re-simulation method, is simply not to use it and do
the high-resolution simulation of the entire initial simulation box.
In \citet[][hereafter Paper I]{Ishiyama2008}, we simulated a
relatively small region (the size of the simulation box is 21.4 Mpc)
with high-mass resolution (mass of particles = $3 \times
10^6M_{\odot}$).  We identified 21 dark halos with maximum rotation
velocity larger than 200$\rm kms^{-1}$ and investigated the
environmental effect on the subhalo abundance. The main conclusion of
Paper I is that the abundance of subhalos has large variation. The most
subhalo-rich halos contain the comparable number of subhalos as
reported in \citet{Moore1999a}, while least subhalo-rich ones contain
around 1/10 subhalos. They also found that only the galaxy-sized halos
(mass less than $3\times 10^{12}M_{\odot}$) show large variation in
subhalo abundance, and more massive ones show much smaller variation.
The subhalo abundance shows correlation with the concentration
parameter $c$ (larger $c$, smaller abundance) and measures of the
local density such as the smoothed local density and distance to the
nearest massive halo.

The accuracy of the result of Paper I is, however, limited by both the
mass resolution (particle mass $3 \times 10^6M_{\odot}$) and small
number statistics. The total number of halos we analyzed was 21, and
only 10 had the mass less than $3 \times 10^{12}M_{\odot}$.  In this
paper, we report the result of much larger, and thus much more
accurate, simulation than that in Paper I.  We simulated a box of the
size 46.48 Mpc with $1600^3$ particles of mass of $1\times
10^{6}M_{\odot}$. We also reduced the softening parameter from 1.8kpc
to 700 pc. Thus, compared to the calculation in Paper I, we used three
times better mass resolution, two times better spatial resolution, and
10 times larger simulation volume.  We identified all halos with the
maximum rotation velocity larger than 160$\rm kms^{-1}$ and the virial
mass $M$ in the range of $1.5\times 10^{12}M_{\odot} \le M < 1\times
10^{13}M_{\odot}$.  We found 125 such halos. In all halos, the number
of particles in the virial radius is more than $1.5\times 10^6$, which
is large enough to give reliable results for the statistics of
subhalos with rotation velocity more than 1/10 of that of the parent
halo \citep{Kase2007}.

In Section 2, we describe the initial model and the simulation
method. In Section 3, we present the result. Section 4 presents summary
and discussion.

\section{Initial model and Numerical Method}

We adopted a LCDM ($\Omega_0=0.3$, $\lambda_0=0.7$, $h=0.7$,
$\sigma_8=0.8$) cosmological model for all of our calculations. In the
largest simulation (simulation I), we used a cube of the comoving size
of 46.48Mpc and periodic boundary condition as the simulation box. The
number of particles is $1600^3$ and the mass of one particle is $10^6
M_{\odot}$.  To generate initial particle distributions, we used the
MPGRAFIC package \citep{Prunet2008}, which is a parallelized variation
of the GRAFIC package \citep{Bertschinger2001}.  The initial redshift was
65.

For the time integration, we used the GreeM code (T.Ishiyama et al. in
preparation), which is a massively parallel TreePM code based on the
parallel TreePM code of \citet{Yoshikawa2005}. It uses the
recursive multisection algorithm \citep{Makino2004} modified to achieve
better load balancing for domain decomposition and the Phantom GRAPE
special force calculation code \citep{Nitadori2006} modified for the
force with cutoff (Nitadori \& Yoshikawa in preparation).  The number
of the grid points for PM calculation was $400^3$.  The opening parameter of
the tree part was $\theta=0.30$ up to $z=10$, and was $\theta=0.50$
from $z=10$ to $z=0$.

We integrated the system using a leapfrog integrator with shared and
adaptive time steps.  The step size was determined as
$\min(2.0\sqrt{\varepsilon/|\vec{a}_i|},2.0\varepsilon/|\vec{v}_i|)$
(minimum of these two values for all particles).
The total number of time steps was $15,844$.  
The (plummer) gravitational softening $\varepsilon$ was
constant in the comoving coordinate up to $z=10$, and was constant
(700 pc) in the physical coordinate from $z=10$ to $z=0$.

We used 2048 CPU cores on the Cray-XT4 at Center for Computational
Astrophysics, CfCA, of National Astronomical Observatory of Japan.
The calculation time per step was about $60$ s.  The full
run ($15,844$ time steps) was completed in $280$ hr (wallclock time).

In order to investigate the effect of the spatial resolution, we
performed two additional simulations (simulations S1 and S2).  We
used the same initial condition for these two runs and changed the
softening parameter.  In table \ref{tab1}, we summarize the number of
particles $N$, the box size $L(\rm Mpc)$, the softening $\varepsilon(\rm
pc)$, the number of steps $N_{\rm step}$, and the redshift $z_{\rm s}$
after which the softening was constant in the physical coordinate, for
the three simulations.

\begin{table}
\caption{Run parameter\label{tab1}}
\begin{tabular}{lrrr}
\hline\hline
Parameter & Sim I & S1 & S2 \\
\hline
$N$ & $1600^3$ & $800^3$ & $800^3$ \\
$L (\rm Mpc)$ & 46.48 & 23.24 & 23.24\\
$\varepsilon (\rm pc)$ & 700 & 700 & 350 \\
$N_{\rm step}$ & 15844 & 10278 & 20513 \\
$z_{\rm s}$ & 10 & 10 & 21\\
\hline 
\end{tabular}
\end{table}

The method to find halos and subhalos is the same as that used in
Paper I.  Our method is based on the idea of finding all local potential minima.  
The identification of the main halos is done in
the following three steps.  In the first step, we sampled particles 
randomly according to the sampling rate $R_{\rm samp}$,
and calculated their potential. 
In this stage, all sampled particles are candidates for the centers of halos.  
Next, we select the particle with the smallest (most negative) potential and list it as the center of a halo.
We then remove $n_{\rm min}$ neighbor particles of this center of halo particles from
the original list of all particles, and select the potential minimum particle from the remaining particles.
At this time, we again look at $n_{\rm min}$ neighbor particles from the list of originally selected particles,
and if one of the neighbors has smaller potential, we do not add this particle to the list of halos.
However, we remove $n_{\rm min}$ neighbors no matter if the particle is added to the list of halos or not.
We repeat this procedure until there are no remaining particles.
Next, we calculated the halo virial radius $R_{\rm v}$ in which the spherical
overdensity is $178\Omega_0^{0.4}$ times the critical value \citep{Eke1996}
, the halo virial mass $M$ and the maximum rotation velocity
$V_{\rm p}$.  For identifying the main halos, we set $R_{\rm
samp}=0.01$ and $n_{\rm min}=1000$ (this value corresponds to
$1.0\times 10^{11}M_{\odot}$).

The method of finding the subhalos is almost the same as that for parent
halos.  The difference is that we do not reduce the particles.  We
detected all local potential minima inside the halo virial radius
$R_{\rm v}$ as subhalos.  For identifying the subhalos, we set $n_{\rm
min}=100$ (this value corresponds to $1.0\times 10^{8}M_{\odot}$).

\section{Result}

We analyzed all halos with the maximum rotation velocity larger than
160 $\rm kms^{-1}$ and the virial mass $1.5\times 10^{12}M_{\odot} \le
M < 1\times 10^{13}M_{\odot}$.  The rotation velocity is defined as
the maximum value of $v_{\rm cir}(r) = \sqrt{GM(r)/r}$ .  We found 68
galaxy-sized halos ($1.5\times10^{12}M_{\odot} \le M <
3\times10^{12}M_{\odot}$), and 57 giant-galaxy-sized halos
($3\times10^{12}M_{\odot} \le M < 1\times10^{13}M_{\odot}$).

\subsection{Effect of the spatial resolution}

\begin{figure}[t]
\centering
\includegraphics[width=8cm, trim=100 0 100 0,clip]{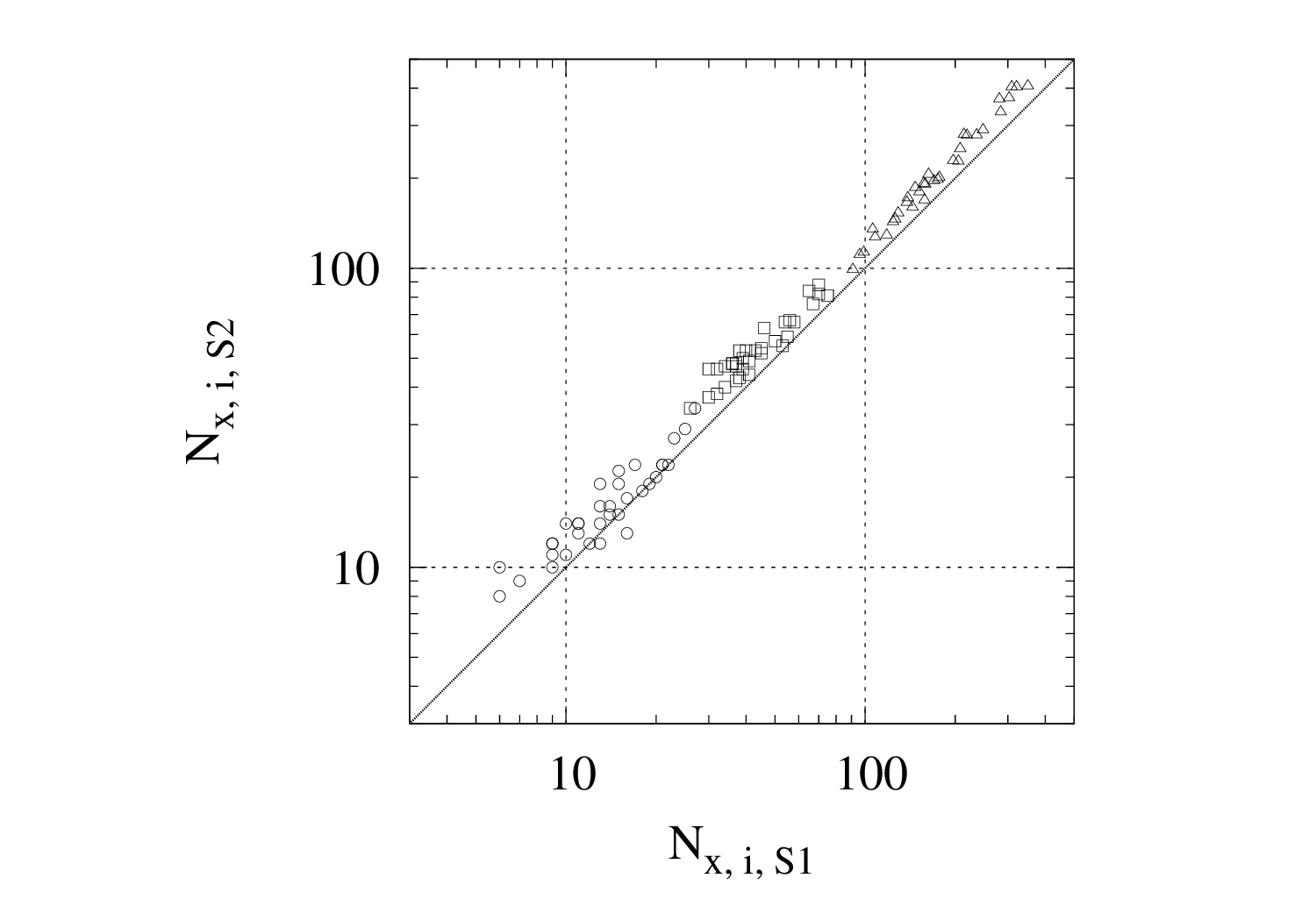}
\caption{Number of subhalos of S1 plotted against those of S2.
Here, $N_{x,i,Sj}$ is the number
of subhalos for halo $i$ with normalized rotation velocity larger
than $x$, in simulation S$j$.
Open circles, squares and triangles show 
$N_{>0.15}$, $N_{>0.10}$ and $N_{>0.05}$, respectively.
}
\label{fig1}
\end{figure}

\begin{table}[h]
\caption{The difference of the subhalo abundance\label{tab2}}
\begin{tabular}{cccc}
\hline\hline
$x$ & 0.15 & 0.10 & 0.05 \\
\hline
$<N_{x, i, S1}>$ & 14.4 & 45.0 & 183.9 \\
$<N_{x, i, S2}>$ & 16.5 & 54.8 & 220.4 \\
$<\Delta N_x/N_x>$ & 0.175 & 0.233 & 0.190 \\
$<(\Delta N_x/N_x)^2>^{1/2}$ & 0.246 & 0.258 & 0.200 \\
\hline 
\end{tabular}
\end{table}

Let us first discuss the reliability of our result. Since the current
calculation used relatively large softening, the difference between
the results of S1 and S2 runs indicates the numerical error of the
main calculation.  We analyzed 21 galaxy-sized halos and 13
giant-galaxy-sized halos in simulations S1 and S2. In these two runs,
we changed only the softening and other parameters are all kept the
same. The reason why we test the effect of the softening here is that
we use the softening significantly larger than that in \citet{Kase2007},
though comparable to what is used in \citet{Springel2008} for
comparable mass resolution (simulation Aq-A-5).

It is known that subhalos contain less than 200 particles is not
reliable \citep{Kase2007}.  The subhalo abundances $N_{>0.15}$,
$N_{>0.10}$ satisfy this condition. However, $N_{>0.05}$ does not.
Here, $N_{>x}$ is defined as the number of subhalos with the
maximum rotation velocity larger than $x$ times that of their parent
halos.

Figure \ref{fig1} shows the subhalo abundance of individual halos in
run S1 plotted against those for the same halos in run S2.  
Open circles, squares, and triangles show  $N_{>0.15}$, $N_{>0.10}$,
and $N_{>0.05}$, respectively.  We can see that halos in run S2 with smaller
softening have more subhalos than their counterparts in run S1. The
difference, however, is small. 

Table \ref{tab2} shows the average relative difference between the number of
subhalos in run S1 and that in run S2, and their standard
deviation. They are calculated as

{\scriptsize
\begin{eqnarray}
<\Delta N_x/N_x> &=& \frac{1}{n_{\rm
halo}}\sum_i \left(\frac{N_{x,i,S2}}{N_{x,i,S1}}-1\right),\\
<(\Delta N_x/N_x)^2>^{1/2} &=& \sqrt{\frac{1}{n_{\rm
halo}}\sum_i \left(\frac{N_{x,i,S2}}{N_{x,i,S1}}-1\right)^2},
\end{eqnarray}
}

where $n_{\rm halo}$ is the number of halos and $N_{x,i,Sj}$ is the number
of subhalos for halo $i$ with normalized rotation velocity larger
than $x$, in simulation S$j$. We can see that the deviation between
the results of runs S1 and S2 is smallest for $x=0.05$.
This is  partly due to small number statistics. As shown in
table \ref{tab2}, the number of subhalos is much smaller for higher rotation
velocity.

The value of $N_{0.05}$ is, however, not reliable because the number
of particles in subhalos with this rotation velocity is too
small. Therefore, we use the value $N_{0.1}$ as the measure of the
subhalo abundance. It does contain the error due to large softening,
but as we have seen in this section that is small. So we ignore this
error in the analysis given in this paper.

\subsection{Subhalo abundance}

\begin{figure}
\centering
\includegraphics[width=8cm]{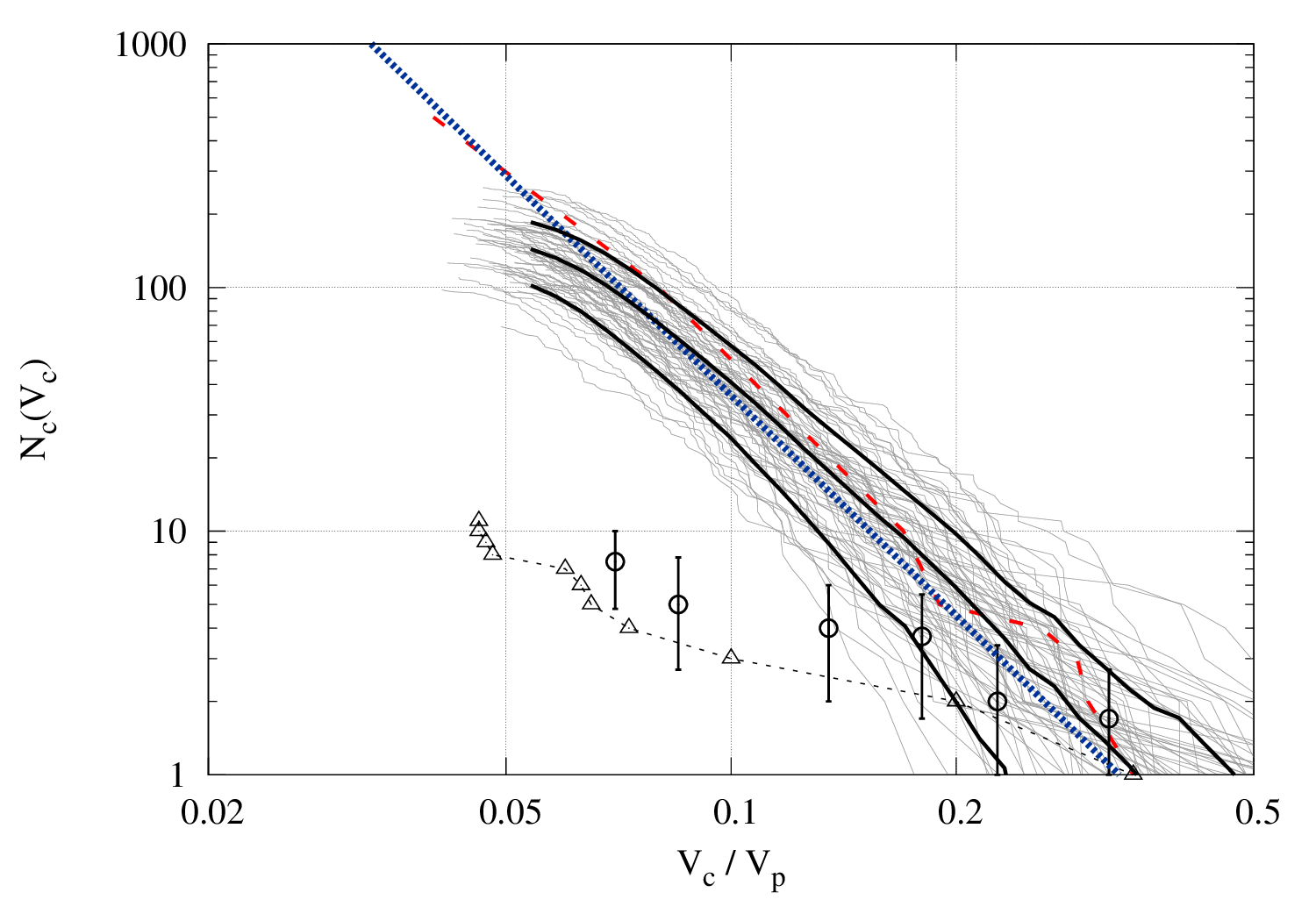}
\includegraphics[width=8cm]{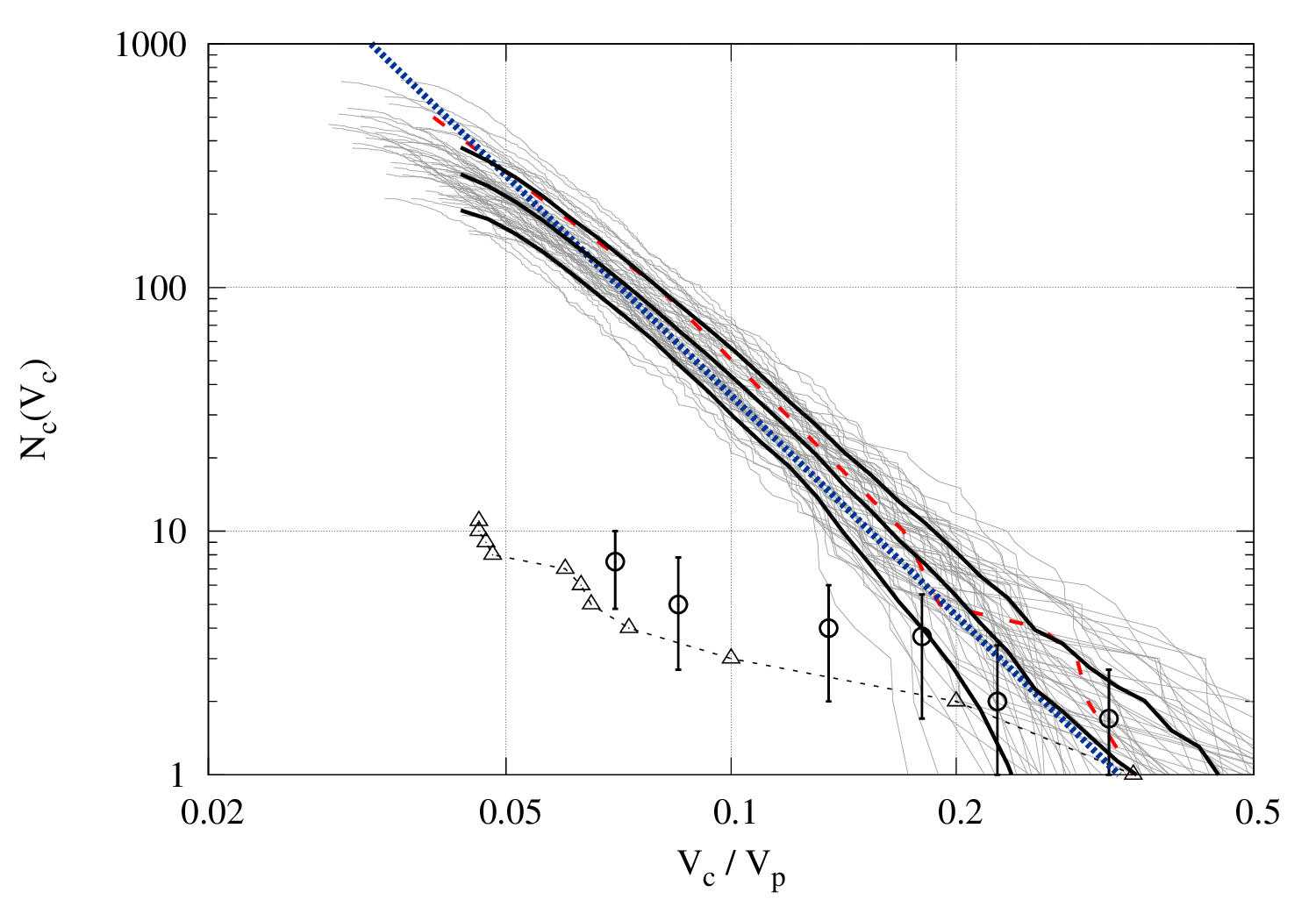}
\caption{
Cumulative numbers of subhalos as a function of their maximum
rotation velocities $V_{\rm c}$ normalized by those of the parent
halos $V_{\rm p}$.  (a) 68 galaxy-sized halos with $1.5\times10^{12}M_{\odot}
\le M < 3\times10^{12}M_{\odot}$ (top).  (b) 57 giant-galaxy-sized halos with
$3\times10^{12}M_{\odot} \le M < 1\times10^{13}M_{\odot}$ (bottom).  
Three thick solid curves show the average (middle) and $\pm 1\sigma$
values (top and bottom). Thick dotted and dashed curves are the
fitting formula from \citet{Diemand2008} and the result of
\citet{Moore1999a} for a galaxy-sized halo, respectively.  The thin
dashed curve with open triangles denotes the number of dwarf galaxies
in our galaxy \citep{Mateo1998}.  The open circles with error bars
show the number of dwarf galaxies in the Local Group
\citep{D'Onghia2007}.
}
\label{fig2}
\end{figure}

Figure \ref{fig2} shows the cumulative numbers of subhalos as a
function of their maximum rotation velocities $V_{\rm c}$ normalized
by those of the parent halos at $z=0$. We can see that the variation
of the subhalo abundance is significantly larger for galaxy-sized
halos (the top panel) than for giant-galaxy-sized halos (the bottom
panel).  The result of \citet{Moore1999a} (thick dashed curve) is
similar to those for our relatively subhalo-rich halos, while that of
\citet{Diemand2008} (thick dotted curve) is comparable to our average
value.  The thin dashed curve with open triangles denotes the dwarf
galaxies in our Galaxy \citep{Mateo1998}, and open circles with error
bars show the dwarf galaxies in the Local Group
\citep{D'Onghia2007}. Note that these observational results do not
include satellites recently found by SDSS \citep{Willman2005,
Belokurov2006, Zucker2006a, Zucker2006b, Belokurov2007, Irwin2007,
Walsh2007}, and thus clearly are underestimates at least for the
rotation velocity less than 10$\rm kms^{-1}$ ($V_{\rm c}/V_{\rm p} <
0.05$).

\begin{figure}[t]
\centering
\includegraphics[width=8cm]{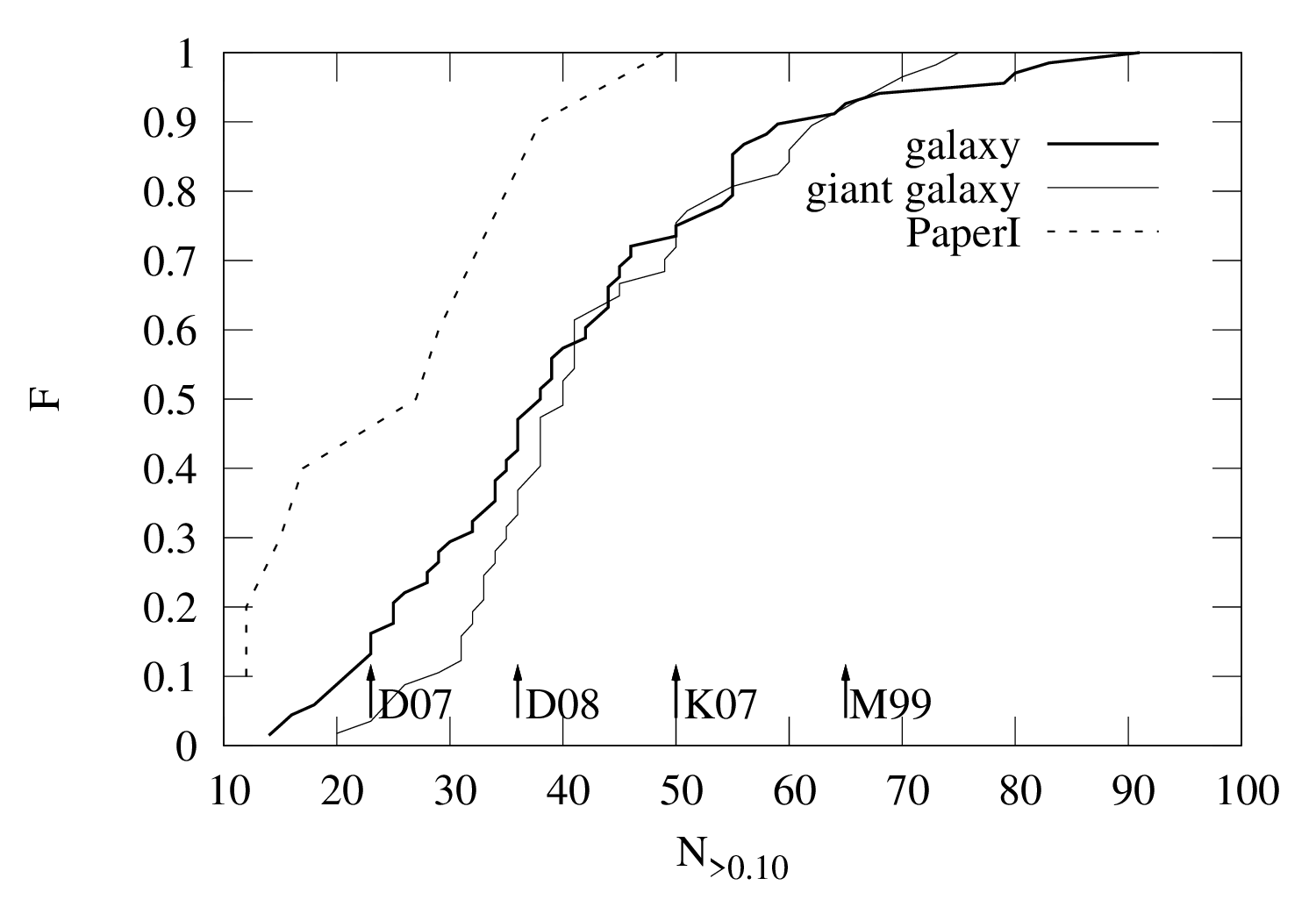}
\caption{
Cumulative function $F$ of parent halos as a function of the number of subhalos with 
the normalized circular velocity larger than 0.1, $N_{>0.1}$.
The thick solid curve shows the result for galaxy-sized halos.
The thin solid curve shows the result for giant galaxy-sized halos.
The thin dashed curve is for the result of Paper I.
The arrows in the bottom show the results of several previous papers
(D07: \citealt{Diemand2007} ; D08: \citealt{Diemand2008}; K07: \citealt{Kase2007}; 
M99: \citealt{Moore1999a}).
}
\label{fig3}
\end{figure}

\begin{figure*}
\centering
\includegraphics[width=8cm]{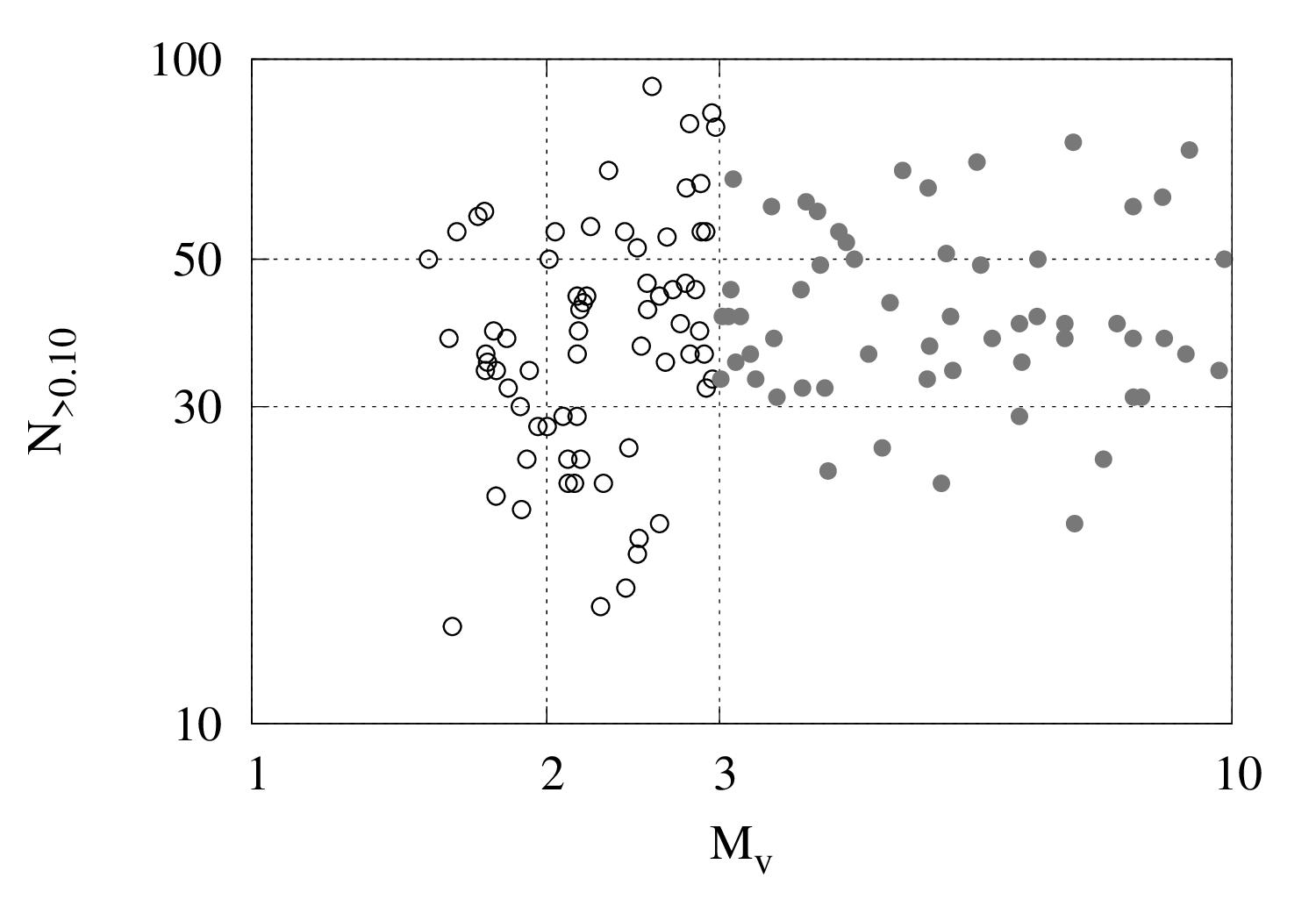}
\includegraphics[width=8cm]{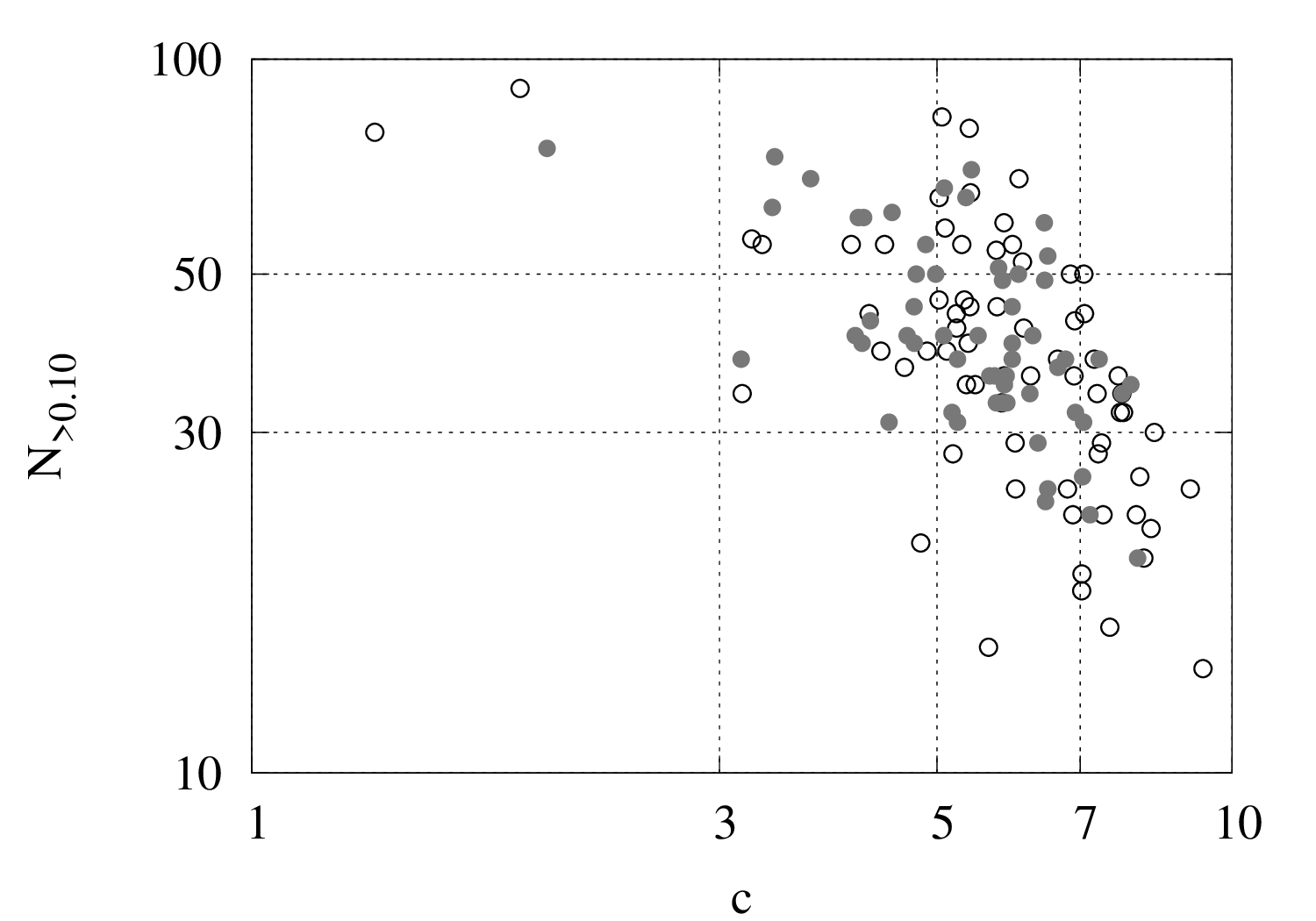}
\caption{
Dependence of the subhalo abundance $N_{>0.10}$ on (a) mass $M_{12}$
in unit of $10^{12}M_{\odot}$ (left), (b) concentration
parameter $c=r_0/R_{\rm v}$ (right).
The white circles denote the halos with
$1.5\times10^{12}M_{\odot} \le M < 3\times10^{12}M_{\odot}$.  The
black circles denote the halos with $3\times10^{12}M_{\odot} \le
M < 1\times10^{13}M_{\odot}$.  }
\label{fig4}
\end{figure*}

Figure \ref{fig3} shows the cumulative distribution of the subhalo
abundance of the parent halos. We used the number of subhalos with the
normalized circular velocity larger than 0.1, $N_{>0.1}$, as the
measure of the subhalo abundance.  We can see that the variation of
the subhalo abundance of galaxy-sized halo is  larger than that
for giant-galaxy-sized halos.  The abundances of the richest halos are
roughly the same for both galaxy-sized and giant-galaxy-sized halos,
but the distribution of galaxy-sized halos is extended to smaller
values of $N_{>0.1}$. We also show the result of several previous
papers.  The results of \citet{Kase2007} and \citet{Diemand2008} 
lie in the middle of the distribution of our halos, while the
result of \citet{Moore1999a} corresponds to the richest ones. 
The differences of \citet{Moore1999a}, 
\citet{Kase2007}, and \citet{Diemand2008} are probably partly
due to the difference in the cosmology. \citet{Moore1999a} and
 \citet{Kase2007} adopted standard CDM without $\Lambda$, 
while \citet{Diemand2008} used
$\Lambda$CDM. Our present calculation and \citet{Diemand2008}  used
quite similar $\Lambda$CDM cosmological parameters. We will return to
the question why the result of  \citet{Diemand2007} corresponds to our
most subhalo-poor halos in Section 4.1.

The dashed curve shows the result of Paper I, for galaxy-sized
halos. The halos Paper I have somewhat smaller number of subhalos, but
since there are only 10 galaxy-sized halos, the difference between
Paper I result and the present result is well within the statistical
variation (not significant for the  KS test with 5\% criterion).

\subsection{The cause of the variation of subhalo abundance}

Figure \ref{fig4}(a) shows the dependence of the parent halo
on the mass.  As already shown in Figure \ref{fig2}, the variation of the
subhalo abundance of galaxy-sized halos is significantly larger than
that for giant-galaxy-sized halos. 

Figure \ref{fig4}(b) shows the dependence on the concentration parameter
$c$. Here,  $c=r_0/R_{\rm v}$, where  $r_0$ is the scale radius of the
fit to the Moore profile \citep{Moore1999b}, and $R_{\rm v}$ is the virial
radius of the halo.  The number of subhalos shows fairly tight
correlation  with the
concentration parameter.  More centrally concentrated halos have fewer
subhalos.  This correlation might mean that the subhalo abundance
depends on the halo formation epoch. Halos with large values of $c$
are generally formed earlier. This difference in the formation time
might be the reason for the difference in the subhalo abundance. 

\begin{figure*}
\centering
\includegraphics[width=3.4cm,angle=270]{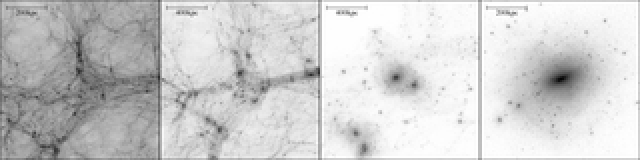}
\includegraphics[width=3.4cm,angle=270]{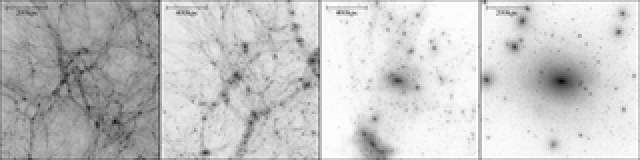}
\includegraphics[width=3.4cm,angle=270]{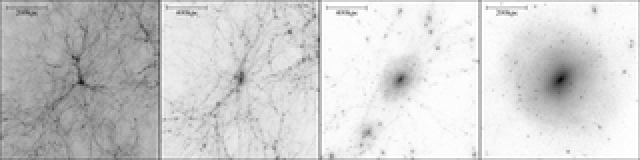}
\includegraphics[width=3.4cm,angle=270]{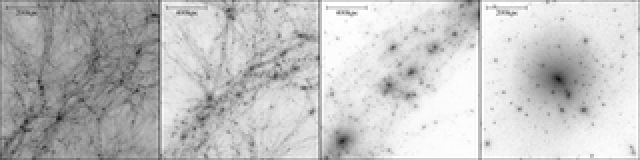}
\includegraphics[width=3.4cm,angle=270]{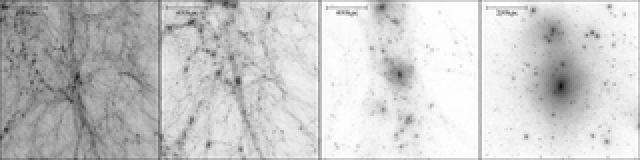}
\includegraphics[width=3.4cm,angle=270]{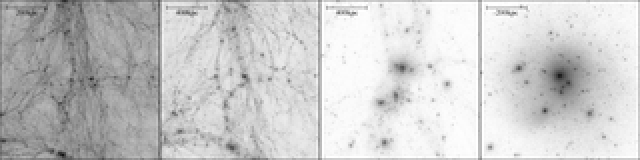}
\caption{
Snapshots of three subhalo-poor halos (top) and three subhalo-rich halos (bottom).
Starting from the left, epoch are $z=6.04$, $z=3.21$, $z=1.04$, and $z=0$,
width are 0.8Mpc, 1.6Mpc, 1.6Mpc, and 0.8Mpc.
The center of snapshot is defined as the center of mass of particles which lie 
in the halo virial radius at $z=0.0$.
}
\label{fig5}
\end{figure*}

\begin{figure*}
\centering
\includegraphics[width=3.4cm,angle=270]{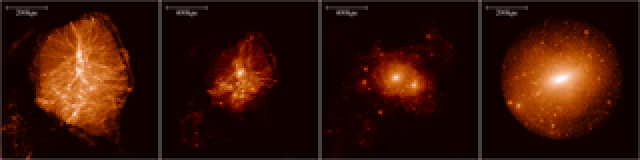}
\includegraphics[width=3.4cm,angle=270]{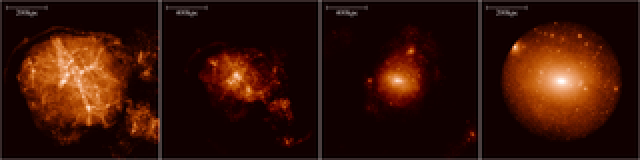}
\includegraphics[width=3.4cm,angle=270]{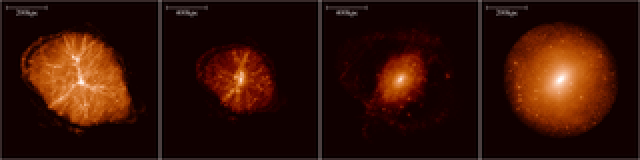}
\includegraphics[width=3.4cm,angle=270]{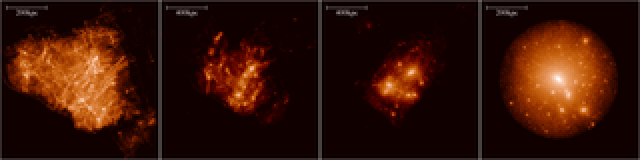}
\includegraphics[width=3.4cm,angle=270]{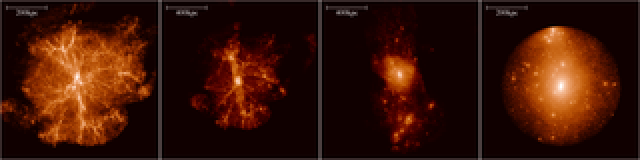}
\includegraphics[width=3.4cm,angle=270]{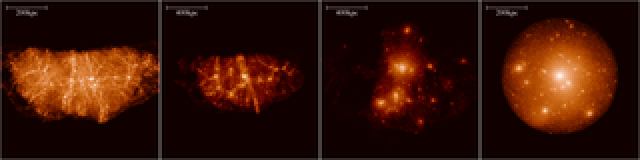}
\caption{
Same as Figure \ref{fig5}. However, 
only particles which lie in the halo virial radius at $z=0$ are plotted.
}
\label{fig6}
\end{figure*}

In Paper I, we found dependences on the local density within 5Mpc
sphere normalized by the average density and also on the distance to the
nearest influential larger-sized halo, which has three times or more
mass of the parent halos.  However, with the improved statistics of
the present paper,  we could not confirm these results.

Figure \ref{fig5} and \ref{fig6} show three subhalo-poor halos (top)
and three subhalo-rich halos (bottom) at $z=6.04$, $z=3.21$, $z=1.04$,
and $z=0$ (left to right).   
They correspond to 0.85, 2.0, 5.68, and 13.7Gyr.
In Figure \ref{fig6}, only particles
which lie in the halo virial radius at $z=0$ are plotted.  As can be
seen from the panels for $z=0$ in Figure \ref{fig6}, there is a large
difference of the subhalo abundance.  The large difference in the
subhalo abundance is already clear at $z=1.04$. At this moment,
subhalo-poor halos have almost finished the assembling (except for the
one in the top row, which still contains two large subhalos). On the
other hand, subhalo-rich halos contain many more large halos which are
in the process of first fallback. Moreover, they are part of much
larger, Mpc-scale filamentary structures. In panels for $z=3.21$,
again, the difference between subhalo-poor halos and rich halos is
rather clear. Rich ones contain many small structures scattered all
over the distribution of particles which form the halo at $z=0$.
In the case of poor halos, at $z=3.21$  single high-density structure
at the center is clearly visible.

When we compare the snapshot at high-$z$ in Figure \ref{fig6}, we can
also see that the physical size of the distribution of particles is
significantly smaller for poor halos, compared to rich halos. Also,
the shape of the halo is smoother and rounder for poor halos.

Thus, from the way these halos are formed, as shown in figures
\ref{fig5} and \ref{fig6}, it is rather clear why there is large
variation in the subhalo abundance. Poor halos are formed from more
centrally concentrated initial condition and formed earlier, while
rich halos are formed from less concentrated structure and formed
later.

In order to quantify the difference in the way the halos are formed,
we tried many different measures, such as the time of the last major
merger and the time at which the mass within the virial radius (or its
certain fraction) exceeds 50\% (or other fraction) of the final
mass. Most measures we tried show the correlation with the subhalo
abundance, but that was weaker than that of the concentration
parameter $c$. One measure with a fair success is the half-mass radius
at the maximum expansion,  $R_{\rm h, max}$, defined as the maximum
value of the radius in which the half of the final virial mass is
enclosed. To calculate this radius, 
we used the potential minimum as the center of the sphere. 

\begin{figure*}
\centering
\includegraphics[width=8cm]{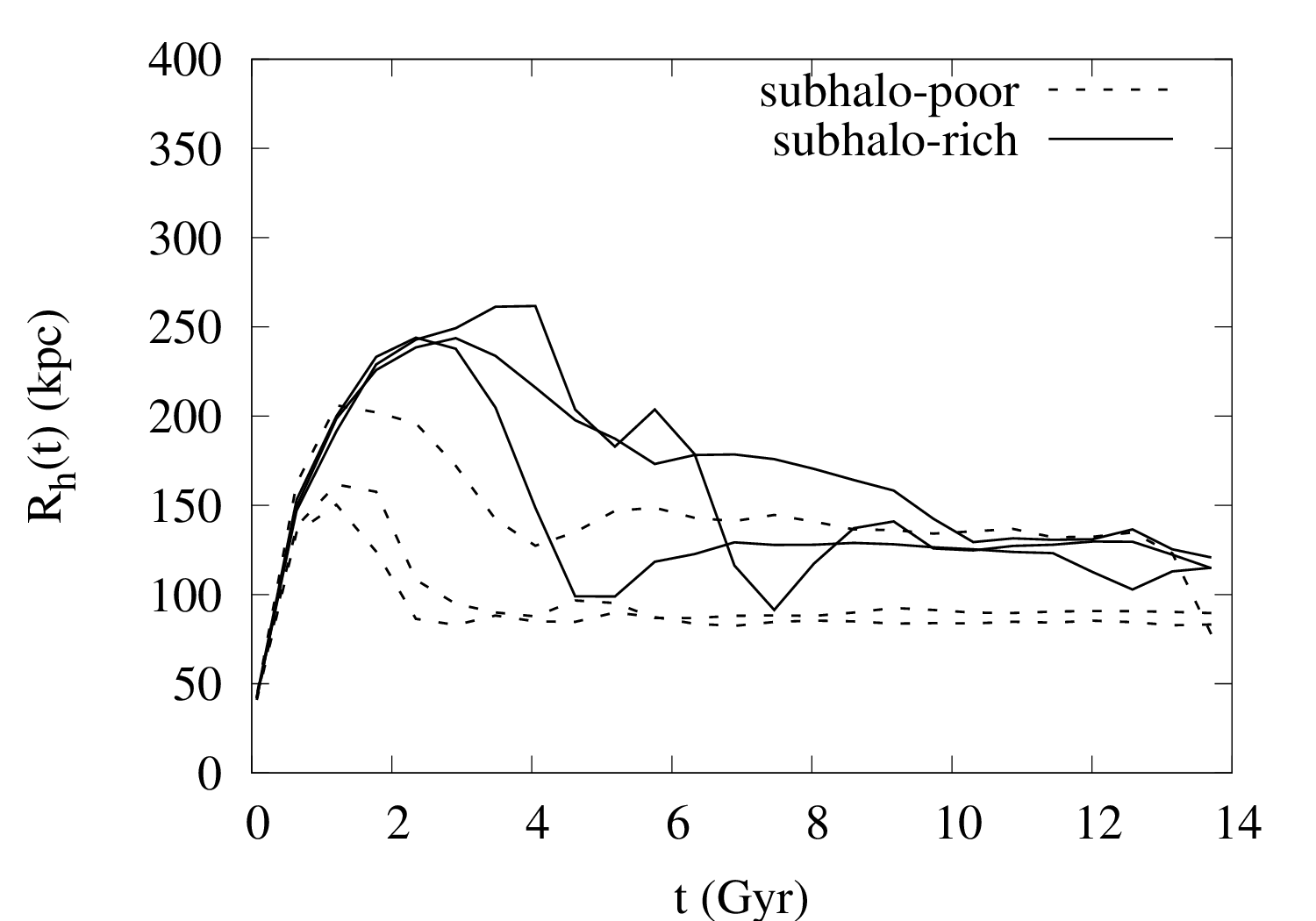}
\includegraphics[width=8cm]{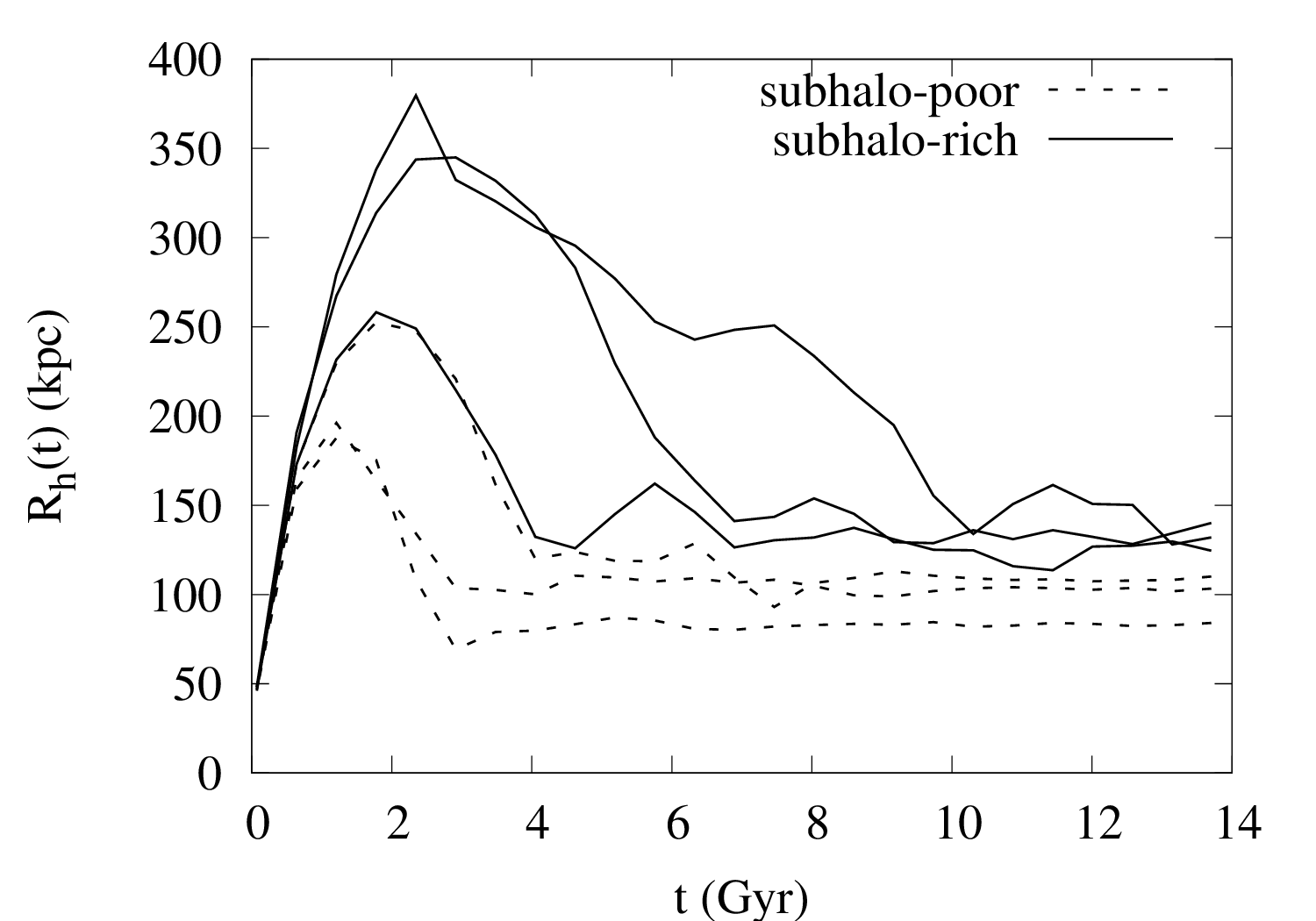}
\caption{
Evolution of the halo half-mass radius $R_{\rm h}(t)$.
The solid curves show the results of three subhalo-rich halos.
The dashed curves show the results of three subhalo-poor halos.
The left panel is for halos with $1.5\times10^{12}M_{\odot} \le M < 2.0\times10^{12}M_{\odot}$.
The right panel is for halos with $2.0\times10^{12}M_{\odot} \le M < 2.5\times10^{12}M_{\odot}$.
}
\label{fig7}
\end{figure*}

Figure \ref{fig7} shows the evolution of the half-mass radius $R_{\rm
h}$.  The solid and dashed curves  show the results of the
rich and poor halos, respectively.

\begin{figure*}
\centering
\includegraphics[width=8cm]{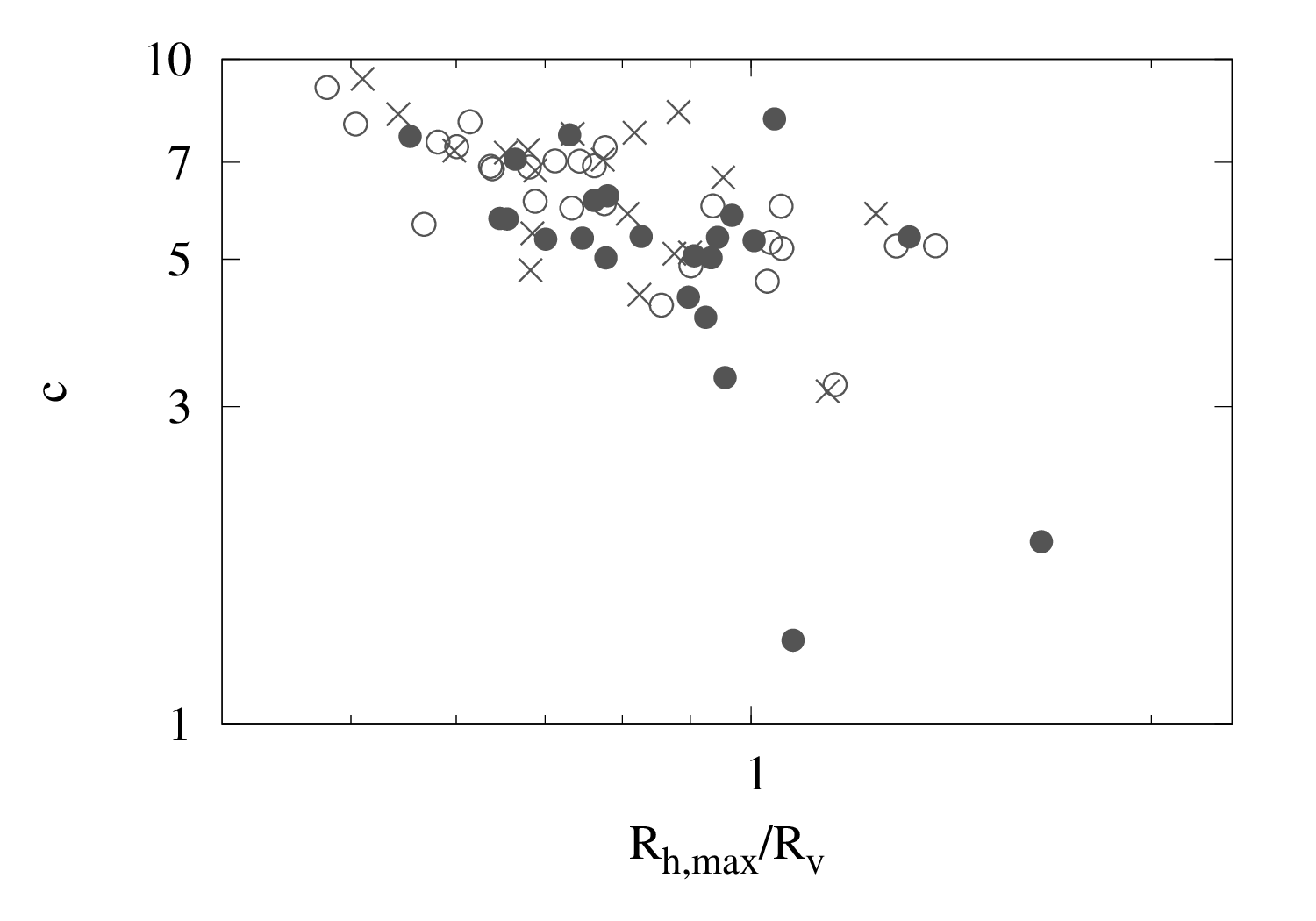}
\includegraphics[width=8cm]{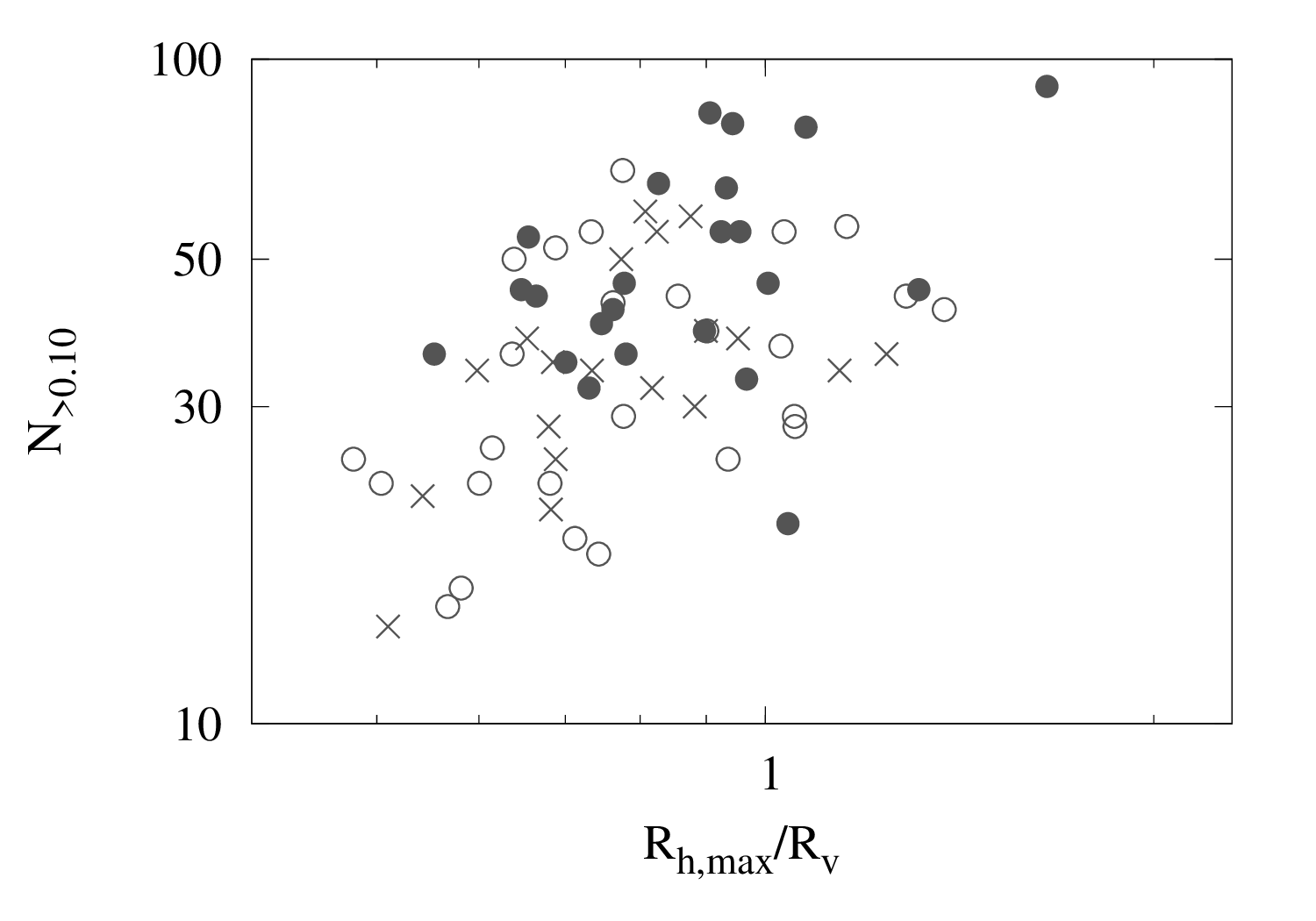}
\caption{
(a) Concentration $c$ plotted against the maximum half-mass radius $R_{\rm h,max}$ 
normalized by the halo virial radius $R_{\rm v}$ (left).
(b) The subhalo abundance $N_{>0.10}$ plotted against the maximum half-mass radius $R_{\rm h,max}$ 
normalized by the halo virial radius $R_{\rm v}$ (right).
Black circles, white circles, and crosses denote
the halos with $2.5\times10^{12}M_{\odot} \le M < 3\times10^{12}M_{\odot}$,
$2.0\times10^{12}M_{\odot} \le M < 2.5\times10^{12}M_{\odot}$, and 
$1.5\times10^{12}M_{\odot} \le M < 2.0\times10^{12}M_{\odot}$, respectively.
}
\label{fig8}
\end{figure*}

\begin{figure*}
\centering
\includegraphics[width=3.4cm,angle=270]{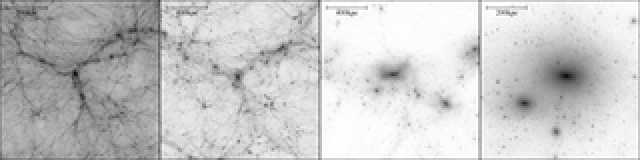}
\includegraphics[width=3.4cm,angle=270]{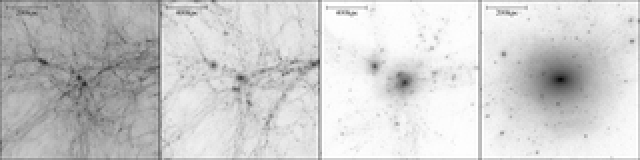}
\includegraphics[width=3.4cm,angle=270]{fig5a.eps}
\includegraphics[width=3.4cm,angle=270]{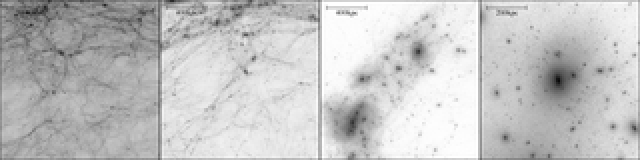}
\includegraphics[width=3.4cm,angle=270]{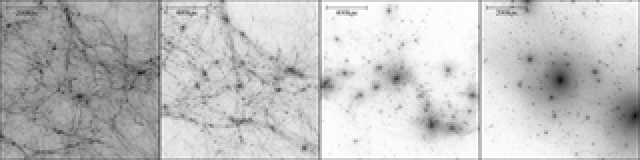}
\includegraphics[width=3.4cm,angle=270]{fig5e.eps}
\caption{
Snapshots of three subhalo-poor halos (top) and three subhalo-rich halos (bottom).
Starting from the left, epoch are $z=6.04$, $z=3.21$, $z=1.04$, and $z=0$,
width are 0.8Mpc, 1.6Mpc, 1.6Mpc, and 0.8Mpc.
The center of snapshot is defined as the center of mass of particles which lie 
in the halo virial radius at $z=0.0$.
}
\label{fig9}
\end{figure*}

\begin{figure*}
\centering
\includegraphics[width=3.4cm,angle=270]{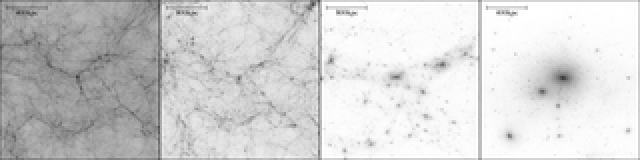}
\includegraphics[width=3.4cm,angle=270]{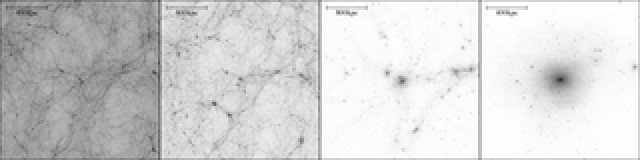}
\includegraphics[width=3.4cm,angle=270]{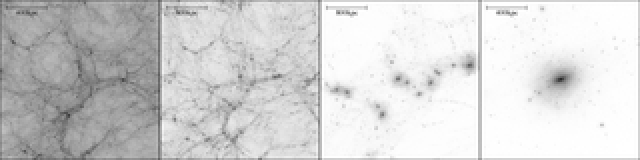}
\includegraphics[width=3.4cm,angle=270]{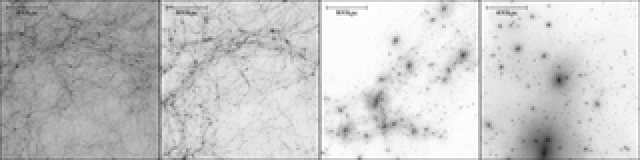}
\includegraphics[width=3.4cm,angle=270]{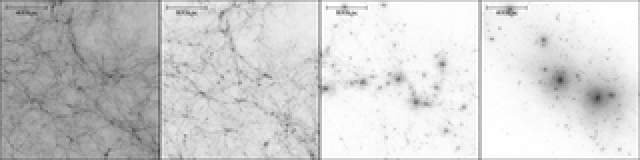}
\includegraphics[width=3.4cm,angle=270]{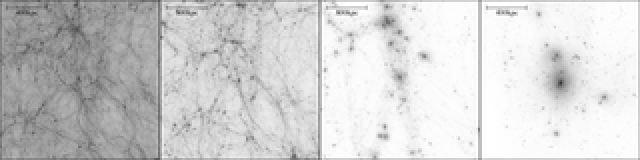}
\caption{
Same as Figure \ref{fig9}. However, 
Starting from the left, width are 1.6Mpc, 3.2Mpc, 3.2Mpc, and 1.6Mpc.
}
\label{fig10}
\end{figure*}

\begin{figure*}
\centering
\includegraphics[width=3.4cm,angle=270]{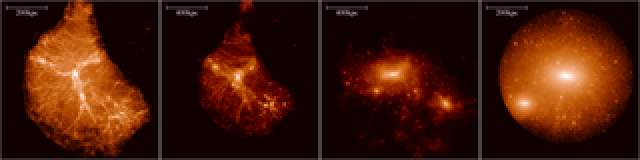}
\includegraphics[width=3.4cm,angle=270]{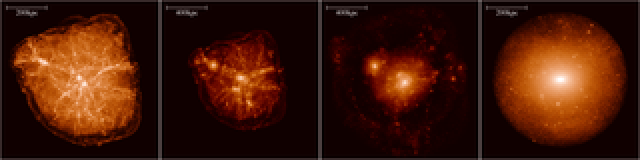}
\includegraphics[width=3.4cm,angle=270]{fig6a.eps}
\includegraphics[width=3.4cm,angle=270]{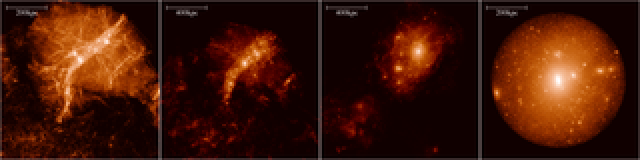}
\includegraphics[width=3.4cm,angle=270]{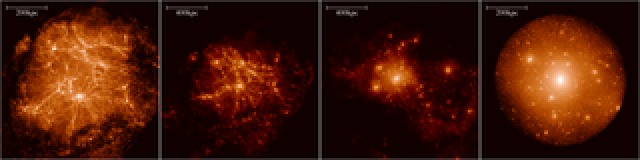}
\includegraphics[width=3.4cm,angle=270]{fig6e.eps}
\caption{
Same as Figure \ref{fig9}. However, 
only particles which lie in the halo virial radius at $z=0$ are plotted.
}
\label{fig11}
\end{figure*}

We can see that $R_{\rm h}$ of poor halos reach the maximum
values much earlier than those for rich halos, and the maximum values
are  smaller for poor halos. Also, 
$R_{\rm h}$ of poor halos seem to settle down faster than 
those of subhalo-rich halos. 

Figure \ref{fig8} shows the relation between the concentration parameter
$c$ and  the maximum half-mass radius $R_{\rm h,max}$ normalized by the halo
virial radius $R_{\rm v}$ (left), and also $N_{>0.1}$ and $R_{\rm
h,max}$. We can see that the correlation is fairly tight for both
cases. Thus, the number of subhalos is at least partly determined by
the way the halo is assembled.

Figure \ref{fig9}-\ref{fig11} show three subhalo-poor halos (top)
and three subhalo-rich halos (bottom) at $z=6.04$, $z=3.21$, $z=1.04$,
and $z=0$ (left to right).
Their $R_{\rm h,max} / R_{\rm v}$ are almost same 
(upper to bottom, 0.71, 0.74, 0.68, 0.69, 0.78, 0.81)
and their concentration parameter are also rather close (7.0, 7.0, 4.8, 6.1, 6.1, 5.9).
Two of them are halos with $1.5\times10^{12}M_{\odot} \le M < 2.0\times10^{12}M_{\odot}$ (third and sixth from top).
Others are halos with $2.0\times10^{12}M_{\odot} \le M < 2.5\times10^{12}M_{\odot}$.
However, we can see the large difference in the subhalo abundance 
from the panels for $z=0$ in Figure \ref{fig11}.
As can be seen from the panels for $z=1.04$, the difference of the shape of the halo is clear.
At this moment, subhalo-poor halos have one or two large subhalos.
By contrast, subhalo-rich halos have many small subhalos.
In panels for high-$z$, we can see the same feature for subhalo-poor and subhalo-rich halos 
as seen in Figure \ref{fig6}.

These mean that subhalo-poor halos with large maximum half-mass radius are formed 
from more centrally concentrated initial condition, formed earlier 
and have a small number of large companions.
In such a halo, the maximum half-mass radius is large because of these companions.
On the other hand, subhalo-rich halos have many small subhalos at $z=1.04$
and many merger events after that.
This might be the reason for the large difference in the subhalo abundance 
of halos with same maximum half-mass radius.

\section{Discussions and Summary}

\subsection{Comparison with Aquarius simulation}

\citet{Springel2008} claimed that they found very small halo-to-halo
variations in the subhalo abundance. They used the definition of the
subhalo count different from what was used in our work and most previous works.
In this section, we compare our result with theirs. 

\begin{figure}
\centering
\includegraphics[width=8cm,trim=80 0 100 0,clip]{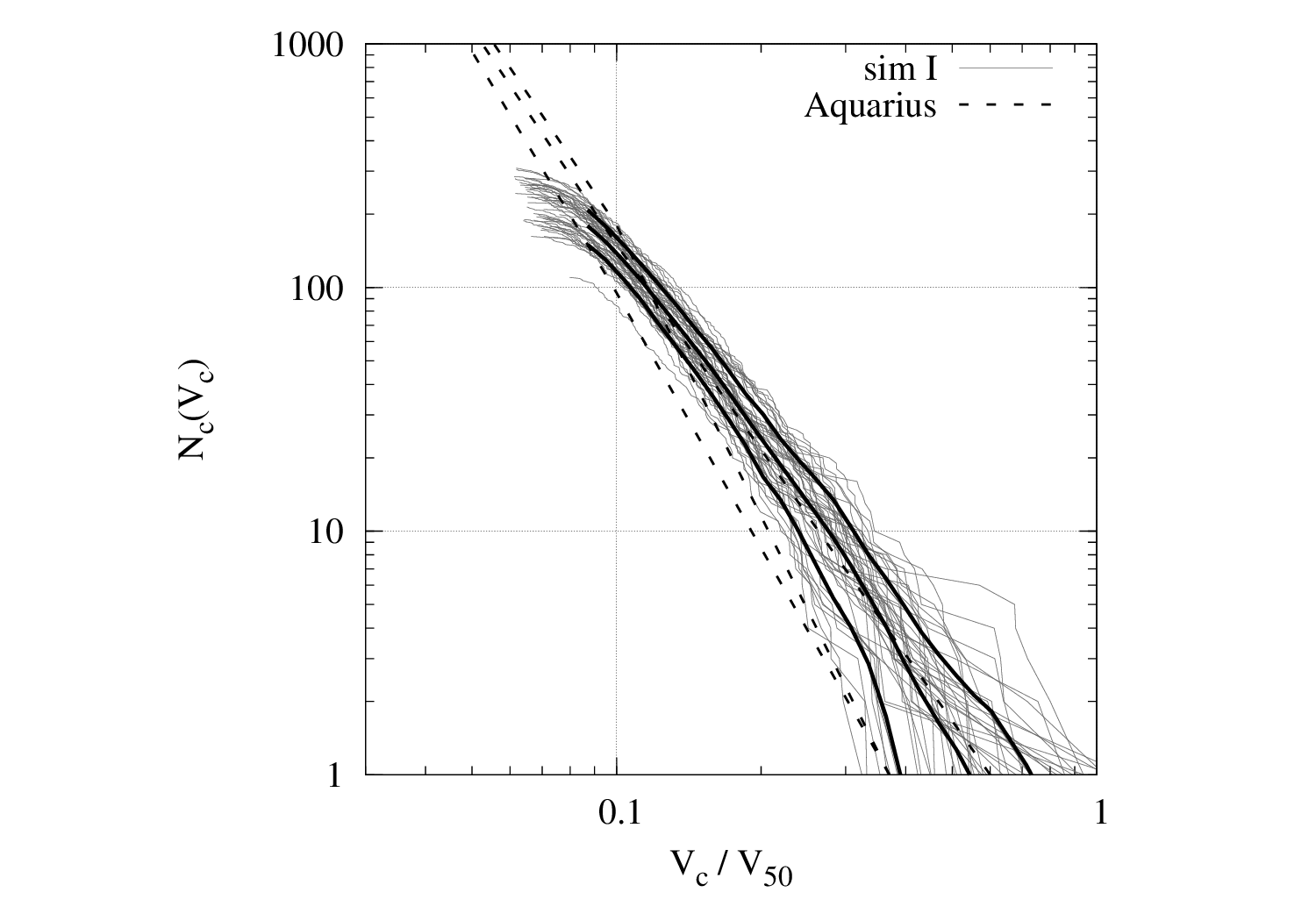}
\caption{
Cumulative numbers of subhalos as a function of their maximum
rotation velocities $V_{\rm c}$ normalized by 
the circular velocity of the parent halos at $r_{50}$.
Three thick solid curves show the average (middle) and $\pm 1\sigma$
values (top and bottom). 
The thick dashed curves show the results of 
\citet[][simulation Aq-A-1, Aq-C-2, Aq-E-2]{Springel2008}.
}
\label{fig12}
\end{figure}

Figure \ref{fig12} shows the cumulative numbers of subhalos as a
function of their maximum rotation velocities $V_{\rm c}$ normalized
by the circular velocity of the parent halos at $r_{50}$.  We define
the radius $r_{50}$ as inside which the averaged spherical overdensity is 50 times the
critical value and count the number of subhalos within $r_{50}$.
This definition is the same as that used for Figure 10 in \citet{Springel2008}.  
We plot the result of the 44 galaxy-sized halos with
$M_{50} < 3.0\times10^{12}M_{\odot}$, where $M_{50}$ is the mass within
$r_{50}$.  

If we measure the subhalo abundance again using $N_{>0.1}$, the number
of subhalos with normalized rotation velocity more than 0.1, we can
conclude that our result and \citet{Springel2008} result are practically
identical. However, our results for $N_{>0.1}$ may be affected by the
resolution, because $V_{50}$ is typically 20\% smaller than the
maximum rotation velocity $V_{\rm p}$.  Thus, our halos are probably
slightly more subhalo-rich than those of \citet{Springel2008}, but the
halo-to-halo variation, calculated using the definition of \citet{Springel2008}, 
is largely similar, though it is difficult to tell much on the
variation in \citet{Springel2008} result because of small number of
runs. We show three results here, and the total number of runs
performed was six.

Compared with Figure \ref{fig2}, the variation of the subhalo
abundance in Figure \ref{fig12} is significantly smaller. This result
is quite natural because of the two differences in the definition of
the subhalo count. First, the use of $r_{50}$ means that many subhalos are
at the outermost region which are not virialized. Though subhalos in
this outer region are gravitationally bound to the main halo, their
evolution is not affected by the structure of the main halo simply
because they are far from the central high-density region. Second, the
use of $V_{50}$ instead of $V_{\rm p}$ would also reduce the halo-to-halo
variation, since centrally concentrated halos have higher $V_{\rm p}$.

\citet{Springel2008} claimed that the small halo-to-halo variation
they found contradicted with the result of Paper I.  Their rms
halo-to-halo scatter was about 8\%, while ours (in Figure 2) is about 41\%.
From the comparison above, it is
clear that this ``contradiction'' they found is partly due to the
difference in the definition of the subhalo abundance.  The
halo-to-halo scatter as seen in Figure \ref{fig12} is smaller than one
in Figure \ref{fig2}.  However, there is still some difference between
our results and their ones.  The remaining difference might be caused by
their way to select parent halos. They selected the halos which host
normal spirals in their semianalytic modelling.

The systematic difference between \citet{Springel2008} results and our
result may be due to the way \citet{Springel2008} selected their halos for
high-resolution calculations. They selected halos without massive
neighbors at $z=0$, and they also put a criterion that in the
semi-analytic calculation the halo should host a late-type
galaxy. This would mean the selected halos should not experience major
merger events at low-$z$, and probably implies that their formation epoch is
early. Thus, it is natural that their halos have systematically
smaller numbers of subhalos than ours have. 
They found systematic difference between their result and 
Via Lactea \citep{Diemand2007,Diemand2008} results.
However, our result agrees with both.

\subsection{Is the missing-dwarf problem solved?}

The difference with the observed number of dwarf galaxies in Local
Group or our galaxy is around a factor of 2 for the most
subhalo-poor ones in our simulation, at the rotation velocity 10\% of
that of the parent halo. For smaller velocities, the difference is
larger, but recent SDSS results indicated that there are still many
dwarf galaxies not identified. Thus, the missing dwarf problem does not
seem to be a very serious problem. A factor-of-two difference may be
easily explained by the effect of baryons. Also, \citet{Springel2008} found
the subhalo abundance even lower than the lowest number we found.
Thus, if our galaxy corresponds to most subhalo-poor halos formed in
$\Lambda$CDM universe, there is little discrepancy between the
observed number of dwarf galaxies and the expected number of dark
matter subhalos.

Obvious question is why our galaxy belongs to most subhalo-poor halos,
but as we found the halos with high central concentration and early
formation epoch have generally smaller number of subhalos. Also, since
our galaxy is a spiral galaxy, it is unlikely that it experienced
recent major merger event. If we look at the Local Group as a whole,
it will experience a major merger in some future, but in this case we
should look at two main subhalos, our galaxy and M31, since their
structure determine the total number of subhalos in the Local Group.

\subsection{Summary}

We simulated $1600^3$ particles with mass of $1.0\times10^6M_{\odot}$
in a 46.48Mpc cubic box to create an unbiased sample of halos with
resolution high enough to determine the abundance of subhalos. We
confirmed the large variation of subhalo abundance we found in Paper I,
but could not confirm the dependence on environmental parameters such as
average density or distance to the nearest massive halo. However, we
found clear dependence on the concentration parameter, which means the
number of subhalos is determined primarily by the initial condition
and formation time.

The ``missing-dwarf problem'' pointed out by \citet{Moore1999a}
turned out to be much less serious, since now the discrepancy is
around a factor of 2. A detailed study of baryon evolution of dwarf
galaxies might be necessary to close up this remaining discrepancy.

\acknowledgements
We are grateful to Takayuki Saitoh for helpful discussions.  
We thank Keigo Nitadori for his technical advice.  Numerical computations were
carried out on Cray XT4 at Center for Computational Astrophysics,
CfCA, of National Astronomical Observatory of Japan.  This research is
partially supported by the Special Coordination Fund for Promoting
Science and Technology (GRAPE-DR project), Ministry of Education,
Culture, Sports, Science and Technology, Japan.

\end{document}